\documentclass[fleqn,twoside]{article}
\usepackage{espcrc2}

\usepackage{epsf,epsfig}
\usepackage[figuresright]{rotating}

\title{CP Violation and the CKM Matrix}

\author{M. Beneke\address[]{Institut f\"ur Theoretische Physik E,
    Sommerfeldstr. 28, 
    RWTH Aachen, D -  52074 Aachen, Germany}%
        \thanks{Talk presented at Lattice 2001, Berlin, August 2001.}}
       
\begin{document}

\begin{abstract}
\noindent
This lecture provides a general overview of CP violation, emphasizing 
CP violation in flavour-violating interactions, such as due to the 
Kobayashi-Maskawa mechanism.
\vspace{1pc}
\end{abstract}

\maketitle

\vspace*{-6.6cm}
\noindent PITHA02/02\\
hep-lat/0201011
\vspace*{4.9cm}

\section{INTRODUCTION}

The breaking of CP symmetry (``CP violation''), the composition of 
parity and charge conjugation, is an interesting phenomenon for 
several reasons:

a. CP violation together with CPT symmetry implies non-invariance of
the microscopic equations of motion under motion-reversal. CP violation 
rather than C violation implies different physical properties of
matter and antimatter. These two facts are of fundamental importance 
for our understanding of the laws of Nature, and they were perceived
as  revolutionary, when CP violation was discovered in 
1964 \cite{Christenson:fg}. They were also important in the
development of the 
fundamental theory of particles, since the observation of CP violation 
motivated some early extensions of the 
Standard Model as it was known at the time (1973), either by extending 
the Higgs sector \cite{Lee:iz} or by adding a third generation of 
quarks and leptons \cite{Kobayashi:fv}. 
Nature has opted for the second possibility  
for certain and the Kobayashi-Maskawa mechanism of CP violation has become 
part of today's Standard Model. From today's perspective
motion-reversal non-invariance and the distinction of matter and 
antimatter, though fundamental, appear no longer surprising and even 
``natural''. What remains surprising, however, is the peculiar way 
in which CP violation occurs or rather does not occur in the Standard Model 
and its possible extensions.

b. Assuming that the evolution of the universe began from a
matter-antimatter symmetric state, CP violation is necessary 
\cite{Sakharov:dj} to 
generate the matter-antimatter asymmetric universe that one observes
today. At the electroweak phase transition the Standard Model 
satisfies all the other necessary criteria for baryogenesis (baryon 
number violation, departure from thermal equilibrium) 
\cite{Kuzmin:1985mm}, but CP
violation in the Standard Model is too weak to explain the observed
baryon-to-photon ratio. With the above assumption on the initial 
condition of the cosmic evolution, our own existence provides evidence
for a source of CP violation beyond the Standard Model.

Electroweak baryogenesis has the attractive feature that it couples 
the required new mechanisms of CP violation to the electroweak scale, 
therefore making them testable also in particle collider experiments. 
Nevertheless it now appears more likely that the matter-antimatter
asymmetry is not related to the sources of CP violation that one may 
observe at colliders. Two facts have contributed to this change of
perspective: first, the lower limit on the masses of Higgs bosons has
been increasing. A heavier Higgs boson implies a weaker (first-order)
electroweak phase transition. As a consequence electroweak
baryogenesis is already too weak over most of the parameter space of
even the minimal supersymmetric extension of the Standard Model. Second, 
the observation of small neutrino masses through neutrino oscillations 
is explained most naturally by invoking the seesaw mechanism, 
which in turn is most naturally realized by postulating
massive neutrinos, which are singlets under the Standard Model gauge
group. All three necessary conditions for the generation of lepton
number are naturally realized in the decay of the massive neutrino(s). 
Lepton number is then partially converted into baryon number via 
B+L-violating (but B--L conserving) sphaleron transitions 
\cite{Fukugita:1986hr}. While the 
leptogenesis scenario is very appealing, the new sources of CP
violation related to the Yukawa couplings of the heavy neutrinos occur 
at scales of order of the heavy neutrino mass, $M_R\sim
(10^{12}-10^{16})\,\mbox{GeV}$ (needed to explain the small left-handed 
neutrino masses), and are not directly testable with collider
experiments in the near future. 
For this reason, CP violation in the context of baryogenesis will not 
be discussed further in this talk.

c. CP violation in the Standard Model is essentially an electroweak 
phenomenon originating from the Yukawa couplings of the quarks to the 
Higgs boson. 
This implies that probes of CP violation are indirect 
probes of the electroweak scale or TeV scale, complementary to direct
probes such as the observation of Higgs bosons. This is probably the 
most important reason for the current interest in CP symmetry 
breaking: in addition to testing the Kobayashi-Maskawa mechanism
of CP violation in the Standard Model, experiments directed at CP
violation limit the construction of extensions of the Standard Model
at the TeV scale. There is an analogy between CP symmetry and 
electroweak symmetry breaking. Both occur at the electroweak scale 
and for both the Standard Model provides a simple mechanism. However,
neither of the two symmetry breaking mechanisms has been sufficiently 
tested up to now. Such tests may or may not confirm the Standard Model
mechanisms but they may also provide answers to questions that concern
the structure of the Standard Model in its entirety, such as the origin of
the electroweak scale and the origin of flavour and CP violation.

d. Leaving aside the matter-antimatter asymmetry in the universe as
evidence for CP violation since this depends on a further assumption, 
CP violation has now been observed in the weak interactions of quarks 
in three different ways: in the mixing of the neutral kaon flavour 
eigenstates ($\epsilon$, 1964) \cite{Christenson:fg}; 
in the decay amplitudes of neutral 
kaons ($\epsilon^\prime/\epsilon$, 1999) 
\cite{Alavi-Harati:1999xp,Fanti:1999nm}; in the mixing of the neutral
$B_d$ meson flavour eigenstates ($\sin(2\beta)$, 2001) 
\cite{Aubert:2001nu,Abe:2001xe}. It will be 
seen below that these pieces of data together with others not directly
related to CP violation suggest that the Kobayashi-Maskawa mechanism
of CP violation is most likely the dominant source of CP violation at
the electroweak scale. The latest piece of evidence also rules out
that CP symmetry is an approximate symmetry. A consequence of this is
that generic extensions of the Standard Model at the TeV scale, needed
to explain the stability of the electroweak scale, suffer from a CP
fine-tuning problem since any such extension implies the existence of
many new CP-violating parameters which have no generic reason to be
small. Despite the apparent success of the standard theory of CP
violation, the problem of CP and flavor violation therefore remains as
mysterious as before. 

The three quantities listed above establish CP violation
unambiguously, but since the observables depend on decays of mesons at
low energy, interpreting these quantities in terms of CP-violating 
fundamental parameters of the theory often involves a very difficult strong
interaction problem. Chiral perturbation theory and the heavy quark
expansion provide analytic tools to address this problem for kaon and 
$B$ meson decays, respectively. In addition, lattice QCD can
contribute substantially to making the theoretical prediction for many
(but not all) quantities relevant to CP violation more precise.

\section{CP VIOLATION IN THE STANDARD MODEL}

CP violation can occur in the Standard Model in three different ways:

\subsection{The $\theta$ term}

The strong interactions could be CP-violating 
\cite{'tHooft:up,Jackiw:1976pf,Callan:je}. The topology of gauge
fields implies that the correct vacuum 
is given by a superposition
$|\theta\rangle = \sum_n e^{i n\theta} |n\rangle$
of the degenerate vacua $|n\rangle$ in which pure gauge fields have 
winding number $n$. Correlation functions in the $\theta$-vacuum can 
be computed by adding to the Lagrangian the term
\begin{equation}
{\cal L}_\theta = \theta\cdot\frac{g_s^2}{32\pi^2}\,G_{\mu\nu}^A 
\tilde G^{A,\mu\nu},
\end{equation}
where $\theta$ now represents a parameter of the theory. Physical
observables can depend on $\theta$ only through the combination 
$e^{i\theta} \det {\cal M}$, where ${\cal M}$ is the quark mass
matrix. A non-zero value of 
\begin{equation}
\tilde \theta =\theta + \mbox{arg}\det {\cal M}
\end{equation}
violates CP symmetry. It also implies an electric dipole moment of 
the neutron of order $10^{-16}\,\tilde \theta \,e\,\mbox{cm}$. The
  non-observation of any such electric dipole moment constrains 
$\tilde \theta < 10^{-10}$ and causes what is known as the strong CP problem, 
since the Standard Model provides no mechanism that would require 
$\tilde \theta$ to vanish naturally. The strong CP problem has become 
more severe with the observation of large CP violation in $B$ meson
decays since one now knows with more confidence that the quark mass 
matrix has no reason to be real a priori. 

There exist mechanisms that render $\tilde\theta=0$ exactly or very 
small through renormalization effects. None of these mechanisms is 
convincing enough to provide a default solution to the problem. 
What makes the strong CP problem so difficult to solve is that one
does not have a clue at what energy scale the solution should be 
sought. Strong CP violation is not discussed further in this talk  
(see the discussion in \cite{Peccei:1998jt}).

\subsection{The neutrino mass matrix}

The Standard Model is an effective theory defined by its gauge 
symmetries and its particle content. CP violation appears in the
lepton sector if neutrinos are massive. The leading operator 
in the effective Lagrangian is \cite{Weinberg:sa}
\begin{equation}
{\displaystyle \frac{f_{ij}}{\Lambda}}
\cdot [(L^T\epsilon)_i i\sigma^2 H][H^T i\sigma^2
L_j].
\end{equation}
After electroweak symmetry breaking this generates a Majorana neutrino
mass matrix with three CP-violating phases. One of these phases could
be observed in neutrino oscillations, the other two phases only in 
observables sensitive to the Majorana nature of neutrinos.

Unless the $f_{ij}$ are extremely small, the scale $\Lambda$ must be
large to account for small neutrinos masses, which suggests that
leptonic CP violation is related to very large scales. For example, 
the standard see-saw mechanism makes the $f_{ij}$ dependent on the 
CP-violating phases in the heavy gauge-singlet neutrino mass matrix. 
As a consequence one may have interesting model-dependent relations 
between leptogenesis, CP violation in lepton-flavour violating
processes and neutrino physics, but since the observations are all
indirect through low-energy experiments, one may at best hope for
accumulating enough evidence to make a particular model particularly
plausible. Such experiments seem to be possible, but not in the near
future and for this reason leptonic CP violation is not discussed
further here. It should be noted that there is in general no connection 
between CP violation in the quark and lepton sector except in grand
unification models in which the two relevant Yukawa matrices are
related. Even then further assumptions are necessary for a
quantitative relation. 

\subsection{The CKM matrix}

CP violation can appear in the quark sector of the Standard Model 
at the level of renormalizable interactions \cite{Kobayashi:fv}. 
The quark Yukawa
interactions read
\begin{equation}
{\cal L}_Y = -y^d_{ij} \bar Q^\prime_i H d^\prime_j - 
y^u_{ij} \bar Q^\prime \epsilon H^*u^\prime_j + {\rm h.c.},
\end{equation}
with $Q^\prime $ the left-handed quark SU(2)-doublets, $u^\prime$ and 
$d^\prime$ the right-handed SU(2) singlets and $i=1,2,3$ the
generation index. The complex mass matrices
that arise after electroweak symmetry breaking are diagonalized by 
separate unitary transformations $U^{u,d}_{L,R}$ of the left- and
right-handed up- and down-type fields. Only the combination 
\begin{equation}
V_{\rm CKM} = {U_L^u}^\dagger U_L^d = 
\left(\begin{array}{ccc}
V_{ud} & V_{us} & V_{ub}\\
V_{cd} & V_{cs} & V_{cb}\\
V_{td} & V_{ts} & V_{tb}
\end{array}\right),
\end{equation}
referred to as the CKM matrix, is observable, 
since the charged current interactions now read
\begin{equation}
-\frac{e}{\sqrt{2}\sin\theta_W} \bar u_i \gamma^\mu [V_{\rm CKM}]_{ij}
d_j W^+_\mu + \mbox{h.c.}.
\end{equation}
Flavour and CP violation in the quark sector 
can occur in the Standard Model only through charged current 
interactions (assuming $\tilde\theta =0$). With three generations of
quarks, the CKM matrix contains one physical CP-violating phase. 
Any CP-violating observable in flavour-violating processes must be
related to this single phase. The verification or, perhaps rather,
falsification of this highly
constrained scenario is the primary goal of many current 
$B$- and $K$-physics experiments. This type of CP violation is
therefore discussed in some detail in later sections.

\begin{figure}[t]
   \vspace{-2.3cm}
   \epsfysize=24cm
   \epsfxsize=16cm
   \hspace*{-1cm}
   \epsffile{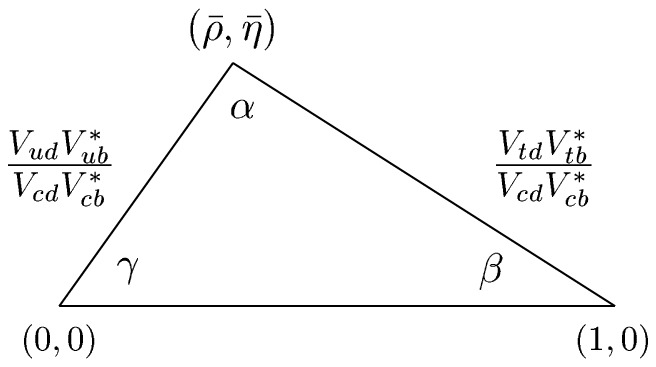}
   \vspace*{-19.6cm}
\caption[dummy]{\label{fig1} The unitarity triangle.}
\end{figure}

For reasons not understood the CKM matrix has a hierarchical structure
as regards transitions between generations. It is therefore often
represented in the approximate form \cite{Wolfenstein:1983yz}
\arraycolsep0.05cm
\begin{equation}
\left(
\begin{array}{ccc}
1-\lambda^2/2 & \lambda & 
A\lambda^3 (\rho-i\eta) \\[0.2cm]
-\lambda & 1-\lambda^2/2 & A\lambda^2 \\[0.2cm]
A\lambda^3 (1-\rho-i\eta) & -A\lambda^2 & 1
\end{array}\right) 
,
\end{equation}
where $\lambda\approx 0.224$ and $A$, $\rho$, $\eta$ are 
counted as order unity and corrections are of order $\lambda^4$. 
It is the great achievement of heavy quark theory of the 1990s
to have determined $|V_{cb}|$, i.e. $A$, to the accuracy of a few
percent, whereas determining $\rho$ and $\eta$ with this accuracy 
remains a 
challenge for this decade. The unitarity of the CKM matrix leads to a
number of relations between rows and columns of the matrix. The one
which is most useful for $B$-physics is obtained by multiplying the
first column by the complex conjugate of the third:
\begin{equation}
V_{ud} V_{ub}^*+V_{cd} V_{cb}^*+V_{td} V_{tb}^*=0.
\end{equation}
If $\eta\not=0$ (which implies CP violation) this relation can be
represented as a triangle in the complex plane, called the unitarity
triangle. See Figure~\ref{fig1}, which also introduces some notation
for the angles of the triangle that will be referred to later on.

The hierarchy of the CKM matrix implies that CP violation is a small
effect in the Standard Model. More precisely, CP-violating observables
are either small numbers, or else they are constructed out of small
numbers such as small branching fractions of rare decays. The
hierarchy of quark masses and mixing angles represents a puzzle, 
sometimes called the flavour problem, which will also not be discussed
further in this talk. 
Typical attempts to solve the flavour problem  
focus on broken generation symmetries.

\section{CONSTRAINTS ON THE UNITARITY TRIANGLE} 

In the following I review the current constraints on $(\bar \rho,
\bar\eta)$, the apex of the unitarity triangle, and the prospects for 
improving these constraints. (The definitions $\bar\rho/\rho = 
\bar\eta/\eta =1 -\lambda^2/2$ render the location of the apex 
accurate to order $\lambda^5$ \cite{Buras:1994ec} 
and will be used in the following.) It is not the purpose of this 
talk to go into the details of the theoretical calculations that 
contribute to these constraints. Recent summaries of the relevant 
lattice calculations can be found in 
\cite{Ryan:2001ej,Martinelli:2001yn,Kronfeld:2001ss}.

\subsection{CP-conserving observables}

The lengths of the three sides of the triangle are determined from 
CP-conserving observables.

{\em Semileptonic decays.} $|V_{cb}|=0.041\pm 0.002$ sets the 
scale of the sides of the triangle and is determined 
from exclusive \cite{Neubert:td} 
and inclusive semileptonic $B$ decays 
\cite{Chay:1990da,Bigi:1992su}. Both methods rely 
on the heavy quark expansion. The current error on $|V_{cb}|$ is not 
a limiting factor in the determination of $(\bar \rho,
\bar\eta)$, but it may become important for rare kaon decays, which 
depend on $A$ to a high power. The inclusive method has probably 
reached its intrinsic limits \cite{Ball:1995wa}. 
Further improvement then depends 
on how well the $B\to D^{(*)}$ form factors can be computed with 
lattice QCD.

The determination of $|V_{ub}|$ uses semileptonic $b\to u$ decays and 
gives
\begin{equation}
\sqrt{\bar \rho^2+\bar\eta^2} = \frac{1-\lambda^2/2}{\lambda}
\left|\frac{V_{ub}}{V_{cb}}\right|.
\end{equation}
However, $|V_{ub}/V_{cb}|\approx 0.085$ is currently known only within 
an error of about $\pm20\%$. $|V_{ub}|$ can also be determined from inclusive 
or exclusive decays. The inclusive treatment would parallel that of 
$|V_{cb}|$ if not the background from $b\to c$ transitions had to be 
suppressed. Distributions in various kinematic variables (lepton energy, 
hadronic invariant mass) have been considered at the price of a more 
complicated and uncertain theory. A cut on the leptonic and hadronic 
invariant mass avoids the kinematic region, where the heavy quark 
expansion (in local operators) is invalid, but the scale of the expansion 
is now around 2$\,$GeV rather than $m_b$ 
\cite{Bauer:2000xf,Neubert:2000ch,Bauer:2001rc}. While the ultimate accuracy 
of this method is not known, it should be possible to halve the error 
on $|V_{ub}|$. The exclusive determination of $|V_{ub}|$ from 
$B\to M l\nu$ must rely on lattice QCD for the $B\to M$ form 
factor. $|V_{ub}|$ is extracted by comparing the lepton invariant mass 
spectrum at large $q^2$ with the form factor in this region, where it is 
computed most reliably at present. The exclusive method is not yet 
competitive, but it will certainly play an important role in an accurate 
determination of $|V_{ub}|$ in the future. 

The $|V_{ub}|$-constraint on $(\bar \rho,\bar\eta)$ is particularly 
important, since it is the only constraint that is based on a tree 
decay and therefore arguably insensitive to non-Standard Model 
interactions. It constrains the parameters of the CKM matrix even 
in the presence of new physics and helps to define the Standard Model 
reference point. The error on $|V_{ub}|$ is also a major component of 
the error in the indirect determination of $\sin 2\beta$ from the 
global fit to  $(\bar \rho,\bar\eta)$ discussed below. Its reduction 
would therefore sharpen the consistency test with the direct 
measurement of $\sin 2\beta$.

{\em $B\bar B$ mixing.} In the Standard Model the $B\bar B$ mass difference is 
dominated by the top quark box diagram, proportional to 
$|V_{tq} V_{tb}^*|^2$ ($q=d,s$). This determines the length of the 
remaining side of unitarity triangle. 
The large mass of the top quark implies that 
the mass difference can be calculated up to the matrix element of 
the local operator $(\bar q b)_{\rm V-A}(\bar q b)_{\rm V-A}$, 
conventionally parameterized by $f_{B_q}^2 B_{B_q}$. For $\Delta M_{B_d}$ 
one obtains
\begin{equation}
\sqrt{(1-\bar\rho)^2+\bar\eta^2} = (0.83\pm 0.03)\times 
\frac{f_{B_d} B_{B_d}^{1/2}}{230\,{\rm MeV}}.
\end{equation}
The use of this result is limited by an error of about 
$\pm 15\%$ on the quantity $f_{B_d} B_{B_d}^{1/2}$. Only lattice QCD 
can possibly improve upon this error. 

Since $V_{ts}$ is already determined by the unitarity of the CKM matrix, 
$\Delta M_{B_s}$ alone does not constrain the unitarity triangle further. 
However, the ratio $(\Delta M_{B_d}/\Delta M_{B_s})^{1/2}$ also 
determines $((1-\bar\rho)^2+\bar\eta^2)^{1/2}$ and involves only the 
ratio $\xi = f_{B_s} B_{B_s}^{1/2}/f_{B_d} B_{B_d}^{1/2}$, which 
is believed to be known from lattice QCD 
with an error ($\pm 6\%$) smaller than the error on each of the 
hadronic parameters individually. The current lower limit on 
$\Delta M_{B_s}$ then provides an important upper limit on the 
length of the relevant side of the unitarity triangle, which turns 
into the most important restriction on the upper limit for the angle 
$\gamma$ in the combined $(\bar \rho,\bar \eta)$ fit.

\subsection{CP violation in kaon decays}

{\em CP violation in mixing (indirect, 1964).} Due to CP violation 
in $K\bar K$ mixing, the neutral kaon mass eigenstates are superpositions 
of CP-even and CP-odd components. The long-lived kaon state, 
$K_L \approx K_2+\bar\epsilon K_1$, is predominantly CP-odd, but decays 
into two pions through its small CP-even component $K_1$. The decay 
$K_L\to\pi\pi$ constituted the first observation of CP violation 
ever \cite{Christenson:fg}. 
The quantity $|\epsilon|=2.27\cdot 10^{-3}$ (equal to $|\bar\epsilon|$ 
in the standard phase convention to very good accuracy) 
has now been measured in many different ways. 

$K\bar K$ mixing is dominated by top and charm box diagrams. The 
long-distance contributions are encoded in the matrix element of 
a four-fermion operator similar to $B\bar B$ mixing, conventionally 
parameterized by $B_K$ and computed in lattice QCD. There is 
currently an uncertainty of at least $\pm 15\%$ on this parameter. 
$|\epsilon|$ determines $\bar\eta \,(1.3\pm 0.05-\bar\rho)$. This is 
the fourth and last constraint that enters the ``standard'' unitarity 
triangle fit.

{\em CP violation in decay (direct, 1999).} CP-violating effects can 
also be seen in the interference of two decay amplitudes with different 
CKM phases. The double ratio
\begin{eqnarray}
&&\frac{\Gamma(K_L\to\pi^0\pi^0)\Gamma(K_S\to\pi^+\pi^-)}
{\Gamma(K_S\to\pi^0\pi^0)\Gamma(K_L\to\pi^+\pi^-)}\nonumber\\
&&\hspace*{3.5cm}
\approx 1-6\mbox{Re}\left(\frac{\epsilon'}{\epsilon}\right),
\end{eqnarray}
if different from unity, implies such an effect, since both ratios 
would equal $|\epsilon|$, if CP violation occurred only in mixing. 
The existence of this effect has been conclusively demonstrated by 
two experiments in 1999, following the first hints of a non-vanishing 
$\epsilon'/\epsilon$ in 1992 \cite{Barr:1993rx}. 
The new results of 2001 have further 
clarified the situation, which is now summarized by 
\cite{Lai:2001ki,ktev2001}
\begin{eqnarray}
{\displaystyle \frac{\epsilon^\prime}{\epsilon}} =
\left\{\begin{array}{ll}
(15.3 \pm 2.6)\cdot 10^{-4} & \quad\mbox{NA48 (97-99)} \\[0.0cm]
(20.7 \pm 2.8)\cdot 10^{-4} & \quad\mbox{KTeV (96/97)}.
\end{array}\right.
\end{eqnarray}

The theory of $\epsilon'/\epsilon$ is more complicated than that of 
any other quantity discussed so far. The short-distance contributions 
are many-fold, but have been worked out to next-to-leading order 
\cite{Ciuchini:1992tj,Buras:1993dy}. 
The following, approximate representation of the result,
\begin{eqnarray}
\frac{\epsilon^\prime}{\epsilon} &=&
16\cdot 10^{-4} \left[{\displaystyle \frac{\mbox{Im}\,V_{ts}^*
      V_{td}}{1.2\cdot 10^{-4}}}\right]
\left({\displaystyle
    \frac{110\,\mbox{MeV}}{m_s(2\,\mbox{GeV})}}\right)^2\\
&& \hspace*{-0.8cm}
\times \left\{\underbrace{B_6^{(1/2)} (1-\Omega_{IB})} 
- \underbrace{0.4 \left({\displaystyle
    \frac{\bar m_t}{165\,\mbox{GeV}}}\right)^{2.5}\!\! \!B_8^{(3/2)}} 
\right\},\nonumber\\
&&\hspace*{0.5cm}\mbox{QCD penguin}\quad\qquad \mbox{
EW penguin}\nonumber
\end{eqnarray}
illustrates the difficulty that arises from a cancellation between 
strong and electroweak penguin contributions and the need to know 
the hadronic matrix elements $B_i \propto 
\langle \pi \pi |O_i| K\rangle$, which involve a two-pion final 
state, accurately. Before 1999 it was commonly, though not universally, 
assumed that $B_{6,8}\approx 1$ near their vacuum saturation value, 
and with the isospin breaking factor $\Omega_{IB}\approx 0.25$, this 
gives only about $6\cdot 10^{-4}$. The experimental result has triggered 
a large theoretical activity directed towards understanding better 
the hadronic matrix elements. Different approaches continue to disagree 
by large factors, but it appears now certain that serious matrix 
element calculations must in one way or another account for final 
state interactions of the two pions. Chiral perturbation theory 
combined with a large-$N_c$ matching of the non-leptonic operators 
can probably go furthest towards this goal with analytic methods. 
The calculation reported in \cite{Pallante:2001he} 
finds $B_6^{(1/2)}$ enhanced by a 
factor 1.55 through rescattering and this, together with a reevaluation 
of isospin-breaking, may account for the experimental result within 
theoretical uncertainties. 

It has also been demonstrated that, in principle, lattice QCD can settle the 
matrix element issue definitively, since $K\to \pi\pi$ matrix 
elements computed in a lattice of finite volume, can be matched 
to continuum, infinite-volume matrix elements including all information 
on rescattering \cite{Lellouch:2000pv,Lin:2001ek}. Due to 
the potential cancellations the matrix elements 
are needed with high precision. They are also needed soon, since 
further CP-violating observables that will be measured in the future 
will diminish the importance of $\epsilon'/\epsilon$.

{\em Rare kaon decays (future).} There exist several proposals to 
measure the very rare decays 
$K^+\to\pi^+\nu\bar\nu$ and $K_L\to\pi^0\nu\bar\nu$ with 
expected branching fractions of about $7\cdot 10^{-11}$ and 
$3\cdot 10^{-11}$, respectively. The first of these decays is 
CP-conserving and constrains $(\bar\rho,\bar\eta)$ to lie on a certain 
ellipse in the $(\bar\rho,\bar\eta)$-plane. The second decay is 
CP-violating and determines $\bar\eta$. The branching fractions 
are predicted theoretically with high precision, so that 
these two kaon modes alone can in principle fix the shape of the 
unitarity triangle, or uncover inconsistencies with other 
constraints \cite{Buchalla:1998ba}. 
Two $K^+\to\pi^+\nu\bar\nu$ events have in fact been 
observed \cite{unknown:2001xv} 
resulting in a branching fraction somewhat larger than 
expected but consistent with expectations within the experimental error.

\subsection{Summing up}

\begin{figure}[t]
   \vspace{0.15cm}
   \epsfysize=7cm
   \epsfxsize=7cm
   \hspace*{0cm}
   \epsffile{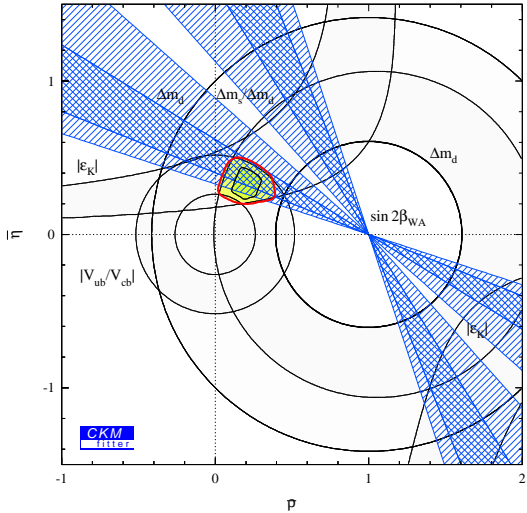}
   \vspace*{-0.7cm}
\caption[dummy]{\label{fig2} Summary of unitarity triangle constraints 
(excluding the direct measurement of $\sin(2\beta)$ which is overlaid)
\cite{Hocker:2001xe}.}
\end{figure}

The four quantities $|V_{ub}/V_{cb}|$, $\Delta M_{B_{d,s}}$ and 
$|\epsilon|$ are usually combined into a global fit of 
$(\bar\rho,\bar\eta)$. Different groups use different statistical 
methods, but since the dominant errors of all input quantities 
are theoretical, no sophisticated procedure can conceal the fact 
that there is a difficulty in quantifying such errors 
objectively. Currently the various procedures appear to 
give similar results when the same inputs are used. 
Figure~\ref{fig2} shows the result of one such global fit 
\cite{Hocker:2001xe}.

The four quantities are in remarkable agreement. This results 
in an indirect determination of the angles of the unitarity triangle,
in particular
\begin{eqnarray}
&&\sin(2\beta)=0.68\pm 0.21,\\
&&\gamma=(58\pm24)^\circ.
\end{eqnarray}
As discussed above the precision of the 
indirect fit relies on the accuracy to which a few hadronic matrix 
elements are known. It is therefore clear that the future of the 
standard unitarity triangle fit (based on the four quantities above) 
is now entirely in the hands of lattice QCD (up to, perhaps, 
$|V_{ub}|$).

\subsection{\boldmath $\sin(2\beta)$}

In 2001 CP violation has been observed also in $B$ meson decays, 
more precisely in the interference of mixing and decay. Assume that 
both, $B^0$ and $\bar B^0$, can decay into a CP eigenstate $f$, 
call the amplitude of the former decay $A$, the latter $\bar A$ and 
define $\lambda=e^{-2 i\beta} \bar A/A$ with $2\beta$ the phase of the 
$B\bar B$ mixing amplitude (standard phase convention). A $B$ meson 
identified as $B^0$ at time $t=0$ can decay into $f$ at a later 
time $t$ either directly or indirectly through its $\bar B^0$
component acquired by mixing. 
If there is CP violation, the amplitude for the 
CP conjugate process will be different, resulting in a time-dependent 
asymmetry
\begin{eqnarray}
A_{\rm CP}(t) &=& \frac{\Gamma(\bar B^0(t)\to f)-\Gamma(B^0(t)\to f)}
{\Gamma(\bar B^0(t)\to f)+\Gamma(B^0(t)\to f)}\\
&&\hspace*{-1.2cm}
= \frac{2{\rm Im}\lambda}{1+|\lambda|^2}\sin(\Delta M_B t)-
\frac{1-|\lambda|^2}{1+|\lambda|^2}\cos(\Delta M_B t).\nonumber
\end{eqnarray}
In the special case that $A$ is dominated by a single weak phase, 
$A=|A| e^{i\delta_W}$ (so that $|\lambda|=1$), the asymmetry is 
proportional to $\pm\sin 2(\beta+\delta_W)$, the sign depending on the 
CP eigenvalue of $f$.

The final state $J/\psi K_S$ (and related ones) satisfies this special 
condition to the accuracy of a percent. Furthermore $\delta_W\approx 0$ 
for $b\to c\bar c s$. Hence the mixing-induced CP asymmetry in 
$B\to J/\psi K$ decay determines the $B\bar B$ mixing phase (relative to 
$b\to c\bar c s$), or $\sin(2\beta)$ in the Standard Model, 
with little theoretical uncertainty
\cite{Bigi:1981qs,Dunietz:1986vi}. 
It determines the $B\bar B$ mixing phase also 
beyond the Standard Model, since it is unlikely that the CKM-favoured 
$b\to c\bar c s$ transition acquires a large CP-violating phase from new 
flavour-changing interactions.

The asymmetry is now precisely measured by the two $B$ factories. The
central values reported by both experiments have been increasing over the 
past year as the statistics of the experiments improved and now 
reads \cite{Aubert:2001nu,Abe:2001xe}
\begin{eqnarray}
\sin(2\beta) =
\left\{\begin{array}{ll}
0.59\pm 0.15 & \quad\mbox{BaBar} \\[0.0cm]
0.99\pm 0.15 & \quad\mbox{Belle},
\end{array}\right.
\end{eqnarray}
yielding the world average $\sin(2\beta)=0.79\pm 0.10$. The fact that 
this asymmetry is large and in agreement with the indirect determination 
of the angle $\beta$ leads to two important conclusions on the 
nature of CP violation:
\begin{itemize}
\item CP is not an approximate symmetry of nature (as could have been 
if CP violation in kaon decays were caused by some non-standard 
interactions).
\item the Kobayashi-Maskawa mechanism of CP violation is most likely 
the dominant source of CP violation at the electroweak scale.
\end{itemize}
New flavour-changing interactions certainly could have affected 
$B\bar B$ mixing and could have been revealed first by the 
direct $\sin(2\beta)$ measurement. If the CKM matrix were the only 
source of flavour-changing processes also in an extension of the 
Standard Model, then the new interactions that modify $B\bar B$ mixing 
also affect $K\bar K$ mixing. One then finds (if one adds an additional 
assumption that there are no new operators in the low-energy effective weak 
Hamiltonian) that only small modifications of the $B\bar B$ mixing 
phase (still related to the phase of $V_{td}$ in such models of 
``minimal flavour violation'') could have been possible, 
in particular $\sin(2\beta)>0.42$ with a preferred range from 
0.5 to 0.8 \cite{Buras:2000xq,Buras:2001af,Bergmann:2001pm,Ali:2001ej}. 
Vice versa the observation of a very small or very large 
$\sin(2\beta)$ would have implied a new mechanism of flavour violation 
with (probably) new CP phases. As discussed below in a more general 
context, this would have led to a CP problem.

\section{MORE CP VIOLATION IN B MESON DECAYS}

The search for CP violation will continue in 
many ways ($B$ decays, $D$ decays, $K$ decays, electric dipole 
moments), but the Kobayashi-Maskawa mechanism predicts large 
effects only in $B$ decays and very rare $K$ decays. The primary 
focus of the coming years will be to verify relations between 
different observables predicted in the Kobayashi-Maskawa scenario 
and to search for small deviations. 

An example of this type is $\bar B_d\to\phi K$ due to 
the penguin $b\to s\bar s s$ transition at the quark level. In 
the Standard Model the time-dependent CP asymmetry of this 
decay is also proportional to $\sin(2\beta)$ to reasonable 
(though not as good) precision. However, new interactions 
are more likely to affect the loop-induced penguin transition 
than the tree decay $b\to c\bar c s$ and may be revealed 
if the time-dependent asymmetry in $\bar B_d\to\phi K$ turns out 
to be different from that in $\bar B_d\to J/\psi K$. However, 
if the difference is small, its interpretation requires that 
one controls the strong interaction effects connected 
with the presence of a small up-quark penguin 
amplitude with a different weak phase. This difficulty is 
of a very general nature in $B$ decays.

\subsection{CP violation in decay}

The need to control strong interaction effects is closely related 
to the possibility of observing CP violation in the decay 
amplitude. The decay amplitude has to have at least two components 
with different weak phases,
\begin{equation}
A(B\to f) = A_1 e^{i\delta_{S1}} e^{i\delta_{W1}}+ 
A_2 e^{i\delta_{S2}} e^{i\delta_{W2}}.
\end{equation}
If the strong interaction phases are also different, the 
partial width of the decay differs from that of its 
CP-conjugate, $\Gamma(B\to f)\not=\Gamma(\bar B\to \bar f)$. 
Many rare $B$ decays are expected to exhibit CP violation 
in decay due to the interference of a tree and a sizeable or 
even dominant penguin amplitude. The weak phase difference 
can only be determined, however, if the strong interaction 
amplitudes are known. This is also necessary for mixing-induced 
CP asymmetries, if the decay amplitude is not dominated by a 
single term.

There exist two complementary approaches to obtain the strong 
interaction amplitudes. The first employs a general 
parameterization of the decay amplitudes of a set of related decays, 
implementing SU(2)-isospin relations. The remaining strong 
interaction parameters are then determined from data (often 
also using SU(3) flavour symmetry and ``little'' further 
assumptions on the magnitudes of some amplitudes). Very often this 
needs difficult measurements. The second approach attempts to 
calculate the strong interaction amplitudes directly from QCD 
with factorization methods also used in high-energy strong interaction 
processes. This approach makes essential use of the fact that 
the $b$ quark mass is large. There is currently no theoretical 
framework that also covers $1/m_b$-corrections systematically, so 
there is an intrinsic limitation to the accuracy that one can 
expect from this approach. Nonetheless, the additional information 
on the dynamics of the decay provided by this approach is 
important as long as data is sparse and will continue to be useful 
later on. 

\subsection{The angle {\boldmath $\gamma$}}

In the Standard Model the angle $\beta$ is obtained accurately 
from the time-dependent CP asymmetry in $B_d\to J/\psi K$. It remains 
to determine directly the angle $\gamma$, the phase of $V_{ub}^*$. 

The preferred methods rely on decays with interference of 
$b\to c\bar u D$ (no phase) and $b\to u\bar c D$ (phase $\gamma$) 
transitions and their conjugates ($D=d,s$). These decays receive no penguin 
contributions and are arguably insensitive to new flavour-changing 
interactions. $\gamma$ can be extracted from either of the 
following decay classes, $B_d(t)\to D^\pm \pi^\mp$ \cite{Dunietz:1987bv}, 
$B_d(t)\to D K_S$ \cite{Gronau:1990ra}, $B^\pm \to K^\pm D_{\rm CP}$ 
\cite{Gronau:1991dp}, $B_s(t)\to D_s^\pm K^\mp$ \cite{Aleksan:1991nh}, 
since every one of them provides sufficiently many observables to eliminate 
all strong interaction parameters, thus providing nice 
illustrations of the first of the two approaches mentioned 
above. None of these strategies is simple to carry out experimentally, 
however, since they involve either small CP asymmetries, or 
small branching fractions, or disparate amplitudes, or 
rapid $B_s$ oscillations.

The possibility to determine $\gamma$ from decays with interference 
of $b\to u\bar u D$ (tree, phase $\gamma$) and $b\to Dq\bar q$ 
(penguin, phase 0 ($D=s$), $\beta$ ($D=d$)) transitions has 
therefore been thoroughly investigated recently, in particular 
the decays $B\to \pi K$. The branching fractions for these modes 
are of order $10^{-5}$ and have already measured with an error of 
$\pm (10-20)\%$, including first measurements of direct CP 
asymmetries (all compatible with zero). The drawback of these and related 
modes is that the amplitudes contain more strong interaction parameters 
than there are 
observables. SU(3) symmetry and the structure of the weak effective 
Hamiltonian allow one to construct a number of interesting bounds 
on $\gamma$ \cite{Fleischer:1998um,Neubert:1998jq}, 
but a full understanding of these modes requires a 
calculation of the penguin-to-tree amplitude ratio including its 
strong rescattering phase. In the following I describe very briefly 
one such method. 

\subsection{QCD factorization}

In the heavy quark limit the $b$ quark decays into very energetic 
quarks (and gluons), 
which must recombine to form two mesons. Methods from the 
heavy quark expansion and soft-collinear factorization 
(``colour-transparency'') can the be argued to imply a factorized 
form of the amplitude of a decay into two light mesons 
\cite{Beneke:1999br,Beneke:2000ry}. 
Schematically,
\begin{eqnarray}
A(\bar B\to M_1 M_2) &=& F^{B\to M_1}(0)\int_0^1 \!\!du\,T^I(u)
\Phi_{M_2}(u) \nonumber\\[0.0cm]
&&\hspace*{-2.8cm}
+\!\int \!\!d\xi du dv T^{II}(\xi,u,v)\,\Phi_B(\xi)\Phi_{M_1}(v) 
\Phi_{M_2}(u),
\end{eqnarray}
where $F^{B\to M_1}$ is a form factor, $\Phi_X$ denote light-cone 
distribution amplitudes and $T^{I,II}$ are perturbative hard-scattering 
kernels, which also contain the strong rescattering phases. This 
result is valid up to $1/m_b$ corrections, some of which can be 
large. The extent to which the QCD factorization formalism can be 
of quantitative use is not yet fully known. The approach has been 
successful in explaining the universality of strong-interaction effects 
in class-I $B\to D$+ light meson 
decays and understanding the non-universality 
in the corresponding class-II decays \cite{Beneke:2000ry}. 
It also appears to account 
naturally for the magnitude of the $\pi K$ branching fractions, 
sometimes considered as unexpectedly large, but there is currently 
no test that would allow one to conclude that the computation of 
strong interaction phases which are either of order $\alpha_s$ or 
$1/m_b$ is reliable in the case of penguin-dominated final states 
\cite{Beneke:2001ev}. 
Such tests will be possible soon and the non-observation of direct 
CP violation at the current level of sensitivity already supports 
the idea that strong rescattering effects are suppressed.  

\begin{figure}[t]
\vspace*{0.2cm}
\epsfxsize=8cm
\centerline{\epsffile{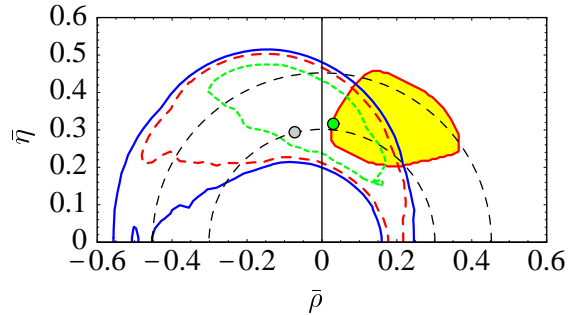}}
\vspace*{-0.6cm}
\caption{\label{fig3}
95\% (solid), 90\% (dashed) and 68\% (short-dashed) confidence level 
contours in the $(\bar\rho,\bar\eta)$ plane obtained from a global 
fit to the CP averaged $B\to\pi K,\pi\pi$ branching fractions, using 
the scanning method \cite{Hocker:2001xe}. The darker dot shows the
overall best fit, whereas the light dot indicates the best fit for the
default parameter set. The light-shaded region indicates the region 
preferred by the standard global fit, excluding the direct measurement 
of $\sin(2\beta)$.}
\end{figure}

Figure~\ref{fig3} shows the result of a global fit of $(\bar \rho,
\bar\eta)$ to CP-averaged $B\to\pi K,\pi\pi$ branching fractions 
with a QCD factorization computation used as an input 
\cite{Beneke:2001ev}. The result is 
consistent with the standard fit based on meson mixing and $V_{ub}$, 
but shows a preference for larger $\gamma$ or smaller $V_{ub}$. 
If the estimate of the theory uncertainty (included in the curves 
in the Figure) is correct, non-leptonic decays together with 
$|V_{ub}|$ from semileptonic decays already imply the existence 
of a CP-violating phase of $V_{ub}$. 

\subsection{Resum\'{e}}

The $B$ factories are already providing data on dozens of rare 
$B$ decay modes. QCD calculations -- 
though probably not very precise -- will be necessary to interpret 
these data beyond ``simple'' quantities like the mixing-induced 
CP asymmetry in $B_d\to J/\psi K$. The immediate future should 
be very interesting since the measurements of direct CP asymmetries 
at the few percent level and the mixing-induced CP asymmetry in 
$B_d\to \pi^+\pi^-$ decay provide tests of the theoretical framework 
and further information on CP violation. Subsequent second generation 
$B$ physics experiments will probably supply enough data to rely more 
and more on measurements and symmetries. Altogether the 
Kobayashi-Maskawa mechanism will be decisively and precisely tested, 
but on the way one can expect many discussions on hadronic physics, 
imagined and, perhaps, true new physics signals.

Lattice calculations will continue to play an important role 
by making more precise the standard unitarity triangle fit. They 
could also provide some of the non-perturbative quantities 
(form factors, light-cone-distribution amplitudes) that are 
needed in factorization-based calculations of non-leptonic decay 
amplitudes. It will be much harder for lattice QCD to make an  
impact on non-leptonic, exclusive decays directly, since 
inelastic rescattering dominates final-state interactions and 
there is currently no method that would allow one to compute 
this on the lattice.

\section{CP VIOLATION IN EXTENSIONS OF THE STANDARD MODEL}

The emerging success of the Kobayashi-Maskawa mechanism of 
CP violation is sometimes accompanied by a sentiment of disappointment 
that the Standard Model has not finally given way to a more fundamental 
theory. However, returning to the perspective of the year 1973, 
when the mechanism was conceived, one can hardly feel this way. 
After all, the Kobayashi-Maskawa mechanism predicted a new 
generation of particles on the basis of the tiny and obscure effect 
of CP violation in $K\bar K$ mixing. It then predicted relations between 
CP-violating quantities in $K$, $D$, $B$-physics which a priori 
might be very different. The fact that it has taken nearly 30 years 
to assemble the experimental tools to test this framework does 
not diminish the spectacular fact that once again Nature has realized 
a structure that was concepted from pure reasoning.

Nevertheless several arguments make it plausible that the 
Kobayashi-Maskawa mechanism is not the final word on CP violation. 
The strong and cosmological CP problem (baryogenesis) continue to 
call for an explanation, probably related to high energy scales. 
There may be an aesthetic appeal to realizing the full Poincar\'{e} 
group as a symmetry of the Lagrangian, in which case CP and P 
symmetry breaking must be spontaneous. One of the strongest arguments 
is, however, that the electroweak hierarchy problem seems to 
require an extension of the Standard Model at the TeV scale. 
Generic extensions have more sources of CP violation than the 
CKM matrix. These have not (yet) been seen, suggesting that there 
is some unknown principle that singles out the CKM matrix as 
the dominant source of flavour and CP violation. In the following 
I give a rather colloquial overview of CP violation in generic 
TeV scale extensions of the Standard Model. This is perhaps an 
academic catalogue, but it illustrates how restrictive the 
Kobayashi-Maskawa framework is.

\subsection{Extended Higgs sector}

Extending the Higgs sector by just a second doublet opens many 
new possibilities. The Higgs potential may now contain complex 
couplings, leading to Higgs bosons without definite CP parity, 
to CP violation in charged Higgs interactions, flavour-changing 
neutral currents, and CP violation in flavour-conserving interactions 
such as $t\bar t H$ and electric dipole moments. 
The Lagrangian could also be CP-conserving with CP violation 
occurring spontaneously through a relative phase of the two 
Higgs vacuum expectations values \cite{Lee:iz}. 

Both scenarios without further restrictions already cause too 
much CP violation and flavour-changing neutral currents, so that 
either the Higgs bosons must be very heavy or some special 
structure imposed. For example, discrete symmetries may 
imply that up-type and down-type quarks couple to only one Higgs 
doublet, a restriction known as ``natural 
flavour conservation'' \cite{Glashow:1976nt,Paschos:1976ay}, 
since it forbids flavour-changing neutral currents (and also 
makes spontaneous CP violation impossible with only two 
doublets). With flavour conservation imposed, CP and flavour violation 
occurs through the CKM matrix, but in addition to the usual 
charged currents also in charged Higgs interactions. This is 
usually considered in the context of supersymmetry, since 
extended Higgs models suffer from the same hierarchy problem 
as the Standard Model.

\subsection{Extended gauge sector (left-right symmetry)}

Left-right-symmetric theories with gauge group SU(2)${}_{\rm L}\times$SU(2)$
{}_{\rm R}\times$U(1)${}_{\rm B-L}$ are attractive
\cite{Mohapatra:1974hk}, 
because parity and CP symmetry can be broken spontaneously. The minimal model 
requires already an elaborate Higgs sector (with triplets in addition 
to doublets) and suffers from the hierarchy problem. CP violation 
in the quark sector now occurs through left- and right-handed 
charged currents with their respective CKM matrices. But since 
all CP violation arises through a single phase in a Higgs vacuum 
expectation value there is now a conflict between suppressing 
flavour-changing currents and having this phase large enough to 
generate the CP violating phenomena already observed. The minimal 
left-right symmetric model with spontaneous CP violation 
is therefore no longer viable 
\cite{Ball:1999mb,Barenboim:2001vu}.

\subsection{Extended fermion sector}

The Standard model can be extended by an extra $d$-type quark 
with electric charge $-1/3$ \cite{delAguila:1985mk}. 
This quark should be a weak singlet in 
order not to conflict electroweak precision tests. After electroweak 
symmetry breaking, the down-quark mass matrix must be diagonalized by a
unitary $4\times 4$ matrix. The motivation for such an extension of 
the Standard Model may be less clear, in particular as there is 
no symmetry principle that would make the extra singlet 
quark naturally light. 
However, this theory provides an example in which the unitarity triangle 
does no longer close to a triangle, but is extended to 
a quadrangle: 
\begin{eqnarray}
&&V_{ud} V_{ub}^*+V_{cd} V_{cb}^*+\underbrace{V_{td} V_{tb}^*}
\,\,\,+\,\,\,\underbrace{U_{db}}\,\,\,=0\\
&& \hspace*{2.5cm}\approx 8\cdot 10^{-3}\quad {<10^{-3}}\nonumber
\end{eqnarray}
The unitarity triangle ``deficit'' $U_{db}$ also determines the 
strength of tree-level flavour-changing $Z$ boson couplings and 
is currently constrained by $B\bar B$ mixing and rare decays to 
about a tenth of the length of a side of the triangle. (The 
corresponding coupling $U_{ds}$ is constrained much more tightly in 
the kaon system.) This model could in principle still give large 
modifications of $B\bar B$ mixing and non-leptonic $B$ decays, 
including CP asymmetries \cite{Barenboim:2001fd}.

\subsection{Supersymmetry}

The minimal supersymmetric standard model 
\cite{Dimopoulos:1981zb,Sakai:1981gr}
is arguably the most natural 
solution to the electroweak hierarchy problem, but it is not particularly 
natural in its most general form from the point of view of CP violation. 
The Lagrangian including the most general Lagrangian that breaks 
supersymmetry softly contains 44 CP-violating constants of nature. 
(R-parity conservation is assumed.) 
One of them is the usual CKM phase which appears in charged current and 
chargino interactions. Three phases appear in flavour-conserving 
CP observables, 27 in flavour- and CP-violating quark-squark-gluino 
interactions (squark mass matrices and $A$-terms) and 13 in the 
(s)lepton sector. 

The flavour-conserving phases must be small to comply with the 
non-observation of electric dipole moments. An intriguing feature of 
supersymmetry is the existence of CP and flavour violation in 
strong interactions (gluinos, squarks). These interactions can be 
much stronger than the Standard Model weak interactions  
and to suppress them 
to a phenomenologically acceptable level, one has to assume 
that either (some of) the masses of superparticles are rather large, 
or that the squark mass matrices are diagonal in the same basis that 
also diagonalizes the quark mass matrices (alignment) or that 
the squarks have degenerate masses, in which case a generalization 
of the GIM mechanism suppresses flavour-changing couplings 
\cite{Ellis:1981ts,Donoghue:1983mx}. Since 
almost all CP-violating phases of the minimal supersymmetric standard model 
originate from supersymmetry breaking terms, one must understand 
supersymmetry breaking to answer the question why CP and flavour 
violation are so strongly suppressed. There exist mechanisms which can 
naturally realize one or the other of the conditions listed above 
(for example supersymmetry breaking through gauge 
interactions \cite{Giudice:1998bp}), 
but none of the mechanisms is somehow singled out. 

There is currently much activity aiming at constraining the flavour- 
and CP-violating couplings from the many pieces of data that become 
now available. In fact these couplings are so many-fold that the 
CP-violating effects observed in kaon and $B$ meson decays can all 
be ascribed to them at the price of making the consistency of the 
Kobayashi-Maskawa mechanism appear accidental. 
The hope could be that eventually some pattern of 
restrictions on these small couplings is seen that could give a hint 
on the origin of supersymmetry breaking. It is also plausible to 
assume that strong flavour and CP violation is absent (or too small 
to observe) in supersymmetry. Neglecting also the flavour-conserving 
CP-violating effects, the CKM matrix is then the only effect 
of interest. The presence of additional particles with CKM couplings 
still implies modifications of meson mixing and rare decays, 
but these modifications are now much smaller and, in general 
(but excepting rare radiative decays), 
precise theoretical results are needed to disentangle them from 
hadronic uncertainties. 

Whatever the outcome of the search for new CP violation may be, 
it will restrict the options for model building severely. The current 
data point towards a privileged standing of the CKM matrix. However, 
a theoretical rationale for this privileged standing is yet to 
be discovered.

\section{CONCLUSIONS}

I. The (expected) observation of large CP violation 
in $B$ decays together with $\epsilon^\prime/\epsilon$ and 
the consistency of indirect determinations
of the unitarity triangle imply that:  

\begin{itemize}
\item[-] CP is not an approximate 
symmetry of nature -- rather CP violation is rare in the Standard 
Model because of small flavour mixing.

\item[-] the Kobayashi-Maskawa mechanism of CP violation 
in charged currents is {\em probably} the {\em dominant} source of CP 
violation {\em at the electroweak scale}.
\end{itemize}

II. CP and electroweak symmetry breaking provide complementary 
motivations to search for extensions of the Standard Model, 
but: 
\begin{itemize}
\item[-] on the one hand, there exists 
no favoured candidate model for CP violation beyond the Standard Model -- 
rather there is a CP problem in many conventional extensions.
\item[- ]on the other hand, baryogenesis requires CP violation 
beyond the Standard Model, probably decoupled from CP violation 
observable at accelerators.
\end{itemize}

III. The study of CP violation is at a turning point with many new 
experimental capabilities and new theoretical methods to interpret 
non-leptonic decay data. Perhaps the most important result  
of the near future, however, will be to find (or not find) the 
$B_s$ mass difference $\Delta M_{B_s}\approx 17.5\,\mbox{ps}^{-1}$, 
confirming once more the Standard Model paradigm (or to put it 
into serious difficulty).

\section*{ACKNOWLEDGEMENT}

\noindent
I wish to thank the organizers, F.~Jegerlehner and M.~M\"uller-Preussker 
in particular, for their support and endless patience.


\begin{thebibliography}{99}

\bibitem{Christenson:fg}
J.~H.~Christenson, J.~W.~Cronin, V.~L.~Fitch and R.~Turlay,
Phys.\ Rev.\ Lett.\  {\bf 13} (1964) 138.

\bibitem{Lee:iz}
T.~D.~Lee,
Phys.\ Rev.\ D {\bf 8} (1973) 1226.

\bibitem{Kobayashi:fv}
M.~Kobayashi and T.~Maskawa,
Prog.\ Theor.\ Phys.\  {\bf 49} (1973) 652.

\bibitem{Sakharov:dj}
A.~D.~Sakharov,
Pisma Zh.\ Eksp.\ Teor.\ Fiz.\  {\bf 5} (1967) 32
[JETP Lett.\  {\bf 5} (1967) 24].

\bibitem{Kuzmin:1985mm}
V.~A.~Kuzmin, V.~A.~Rubakov and M.~E.~Shaposhnikov,
Phys.\ Lett.\ B {\bf 155} (1985) 36.

\bibitem{Fukugita:1986hr}
M.~Fukugita and T.~Yanagida,
Phys.\ Lett.\ B {\bf 174} (1986) 45.

\bibitem{Alavi-Harati:1999xp}
A.~Alavi-Harati {\it et al.}  [KTeV Collaboration],
Phys.\ Rev.\ Lett.\  {\bf 83} (1999) 22
[hep-ex/9905060].

\bibitem{Fanti:1999nm}
V.~Fanti {\it et al.}  [NA48 Collaboration],
Phys.\ Lett.\ B {\bf 465} (1999) 335
[hep-ex/9909022].

\bibitem{Aubert:2001nu}
B.~Aubert {\it et al.}  [BABAR Collaboration],
Phys.\ Rev.\ Lett.\  {\bf 87} (2001) 091801
[hep-ex/0107013].

\bibitem{Abe:2001xe}
K.~Abe {\it et al.}  [Belle Collaboration],
Phys.\ Rev.\ Lett.\  {\bf 87} (2001) 091802
[hep-ex/0107061].

\bibitem{'tHooft:up}
G.~'t Hooft,
Phys.\ Rev.\ Lett.\  {\bf 37} (1976) 8.

\bibitem{Jackiw:1976pf}
R.~Jackiw and C.~Rebbi,
Phys.\ Rev.\ Lett.\  {\bf 37} (1976) 172.

\bibitem{Callan:je}
C.~G.~Callan, R.~F.~Dashen and D.~J.~Gross,
Phys.\ Lett.\ B {\bf 63} (1976) 334.

\bibitem{Peccei:1998jt}
R.~D.~Peccei,
[hep-ph/9807514].

\bibitem{Weinberg:sa}
S.~Weinberg,
Phys.\ Rev.\ Lett.\  {\bf 43} (1979) 1566.

\bibitem{Wolfenstein:1983yz}
L.~Wolfenstein,
Phys.\ Rev.\ Lett.\  {\bf 51} (1983) 1945.

\bibitem{Buras:1994ec}
A.~J.~Buras, M.~E.~Lautenbacher and G.~Ostermaier,
Phys.\ Rev.\ D {\bf 50} (1994) 3433
[hep-ph/9403384].

\bibitem{Ryan:2001ej}
S.~Ryan, talk at this symposium [hep-lat/0111010].

\bibitem{Martinelli:2001yn}
G.~Martinelli, talk at this symposium [hep-lat/0112011].

\bibitem{Kronfeld:2001ss}
A.~S.~Kronfeld, talk at Heavy Flavours 9, Pasadena, September 2001
[hep-ph/0111376].

\bibitem{Neubert:td}
M.~Neubert,
Phys.\ Lett.\ B {\bf 264} (1991) 455.

\bibitem{Chay:1990da}
J.~Chay, H.~Georgi and B.~Grinstein,
Phys.\ Lett.\ B {\bf 247} (1990) 399.

\bibitem{Bigi:1992su}
I.~I.~Bigi, N.~G.~Uraltsev and A.~I.~Vainshtein,
Phys.\ Lett.\ B {\bf 293} (1992) 430
[Erratum-ibid.\ B {\bf 297} (1992) 477]
[hep-ph/9207214].

\bibitem{Ball:1995wa}
P.~Ball, M.~Beneke and V.~M.~Braun,
Phys.\ Rev.\ D {\bf 52} (1995) 3929
[hep-ph/9503492].

\bibitem{Bauer:2000xf}
C.~W.~Bauer, Z.~Ligeti and M.~E.~Luke,
Phys.\ Lett.\ B {\bf 479} (2000) 395
[hep-ph/0002161].

\bibitem{Neubert:2000ch}
M.~Neubert,
JHEP {\bf 0007} (2000) 022
[hep-ph/0006068].

\bibitem{Bauer:2001rc}
C.~W.~Bauer, Z.~Ligeti and M.~E.~Luke,
Phys.\ Rev.\ D {\bf 64} (2001) 113004
[hep-ph/0107074].

\bibitem{Barr:1993rx}
G.~D.~Barr {\it et al.}  [NA31 Collaboration],
Phys.\ Lett.\ B {\bf 317} (1993) 233.

\bibitem{Lai:2001ki}
A.~Lai {\it et al.}  [NA48 Collaboration],
Eur.\ Phys.\ J.\ C {\bf 22} (2001) 231
[hep-ex/0110019].

\bibitem{ktev2001}
A.~Glazov [KTeV Collaboration], talk presented at KAON 2001, 
Pisa, June 2001.

\bibitem{Ciuchini:1992tj}
M.~Ciuchini, E.~Franco, G.~Martinelli and L.~Reina,
Phys.\ Lett.\ B {\bf 301} (1993) 263
[hep-ph/9212203].

\bibitem{Buras:1993dy}
A.~J.~Buras, M.~Jamin and M.~E.~Lautenbacher,
Nucl.\ Phys.\ B {\bf 408} (1993) 209
[hep-ph/9303284].

\bibitem{Pallante:2001he}
E.~Pallante, A.~Pich and I.~Scimemi,
Nucl.\ Phys.\ B {\bf 617} (2001) 441
[hep-ph/0105011].

\bibitem{Lellouch:2000pv}
L.~Lellouch and M.~L\"uscher,
Commun.\ Math.\ Phys.\  {\bf 219} (2001) 31
[hep-lat/0003023].

\bibitem{Lin:2001ek}
C.~J.~Lin, G.~Martinelli, C.~T.~Sachrajda and M.~Testa,
Nucl.\ Phys.\ B {\bf 619} (2001) 467
[hep-lat/0104006].

\bibitem{Buchalla:1998ba}
G.~Buchalla and A.~J.~Buras,
Nucl.\ Phys.\ B {\bf 548} (1999) 309
[hep-ph/9901288].

\bibitem{unknown:2001xv}
S.~Adler {\it et al.} [E787 Collaboration],
[hep-ex/0111091].

\bibitem{Hocker:2001xe}
A.~H\"ocker, H.~Lacker, S.~Laplace and F.~Le Diberder,
Eur.\ Phys.\ J.\ C {\bf 21}, 225 (2001)
[hep-ph/0104062].

\bibitem{Bigi:1981qs}
I.~I.~Bigi and A.~I.~Sanda,
Nucl.\ Phys.\ B {\bf 193}, 85 (1981).

\bibitem{Dunietz:1986vi}
I.~Dunietz and J.~L.~Rosner,
Phys.\ Rev.\ D {\bf 34}, 1404 (1986).

\bibitem{Buras:2000xq}
A.~J.~Buras and R.~Buras,
Phys.\ Lett.\ B {\bf 501} (2001) 223
[hep-ph/0008273].

\bibitem{Buras:2001af}
A.~J.~Buras and R.~Fleischer,
Phys.\ Rev.\ D {\bf 64} (2001) 115010
[hep-ph/0104238].

\bibitem{Bergmann:2001pm}
S.~Bergmann and G.~Perez,
Phys.\ Rev.\ D {\bf 64} (2001) 115009
[hep-ph/0103299].

\bibitem{Ali:2001ej}
A.~Ali and E.~Lunghi,
Eur.\ Phys.\ J.\ C {\bf 21} (2001) 683
[hep-ph/0105200].

\bibitem{Dunietz:1987bv}
I.~Dunietz and R.~G.~Sachs,
Phys.\ Rev.\ D {\bf 37} (1988) 3186
[Erratum-ibid.\ D {\bf 39} (1988) 3515].

\bibitem{Gronau:1990ra}
M.~Gronau and D.~London.,
Phys.\ Lett.\ B {\bf 253} (1991) 483.

\bibitem{Gronau:1991dp}
M.~Gronau and D.~Wyler,
Phys.\ Lett.\ B {\bf 265} (1991) 172.

\bibitem{Aleksan:1991nh}
R.~Aleksan, I.~Dunietz and B.~Kayser,
Z.\ Phys.\ C {\bf 54} (1992) 653.

\bibitem{Fleischer:1998um}
R.~Fleischer and T.~Mannel,
Phys.\ Rev.\ D {\bf 57}, 2752 (1998)
[hep-ph/9704423].

\bibitem{Neubert:1998jq}
M.~Neubert and J.~L.~Rosner,
Phys.\ Rev.\ Lett.\  {\bf 81}, 5076 (1998)
[hep-ph/9809311].

\bibitem{Beneke:1999br}
M.~Beneke, G.~Buchalla, M.~Neubert and C.~T.~Sachrajda,
Phys.\ Rev.\ Lett.\  {\bf 83}, 1914 (1999)
[hep-ph/9905312].

\bibitem{Beneke:2000ry}
M.~Beneke, G.~Buchalla, M.~Neubert and C.~T.~Sachrajda,
Nucl.\ Phys.\ B {\bf 591}, 313 (2000)
[hep-ph/0006124].

\bibitem{Beneke:2001ev}
M.~Beneke, G.~Buchalla, M.~Neubert and C.~T.~Sachrajda,
Nucl.\ Phys.\ B {\bf 606}, 245 (2001)
[hep-ph/0104110].

\bibitem{Glashow:1976nt}
S.~L.~Glashow and S.~Weinberg,
Phys.\ Rev.\ D {\bf 15} (1977) 1958.

\bibitem{Paschos:1976ay}
E.~A.~Paschos,
Phys.\ Rev.\ D {\bf 15} (1977) 1966.

\bibitem{Mohapatra:1974hk}
R.~N.~Mohapatra and J.~C.~Pati,
Phys.\ Rev.\ D {\bf 11} (1975) 566.

\bibitem{Ball:1999mb}
P.~Ball, J.~M.~Frere and J.~Matias,
Nucl.\ Phys.\ B {\bf 572} (2000) 3
[hep-ph/9910211].

\bibitem{Barenboim:2001vu}
G.~Barenboim, M.~Gorbahn, U.~Nierste and M.~Raidal,
[hep-ph/0107121].

\bibitem{delAguila:1985mk}
F.~del Aguila and J.~Cortes,
Phys.\ Lett.\ B {\bf 156} (1985) 243.

\bibitem{Barenboim:2001fd}
G.~Barenboim, F.~J.~Botella and O.~Vives,
Nucl.\ Phys.\ B {\bf 613} (2001) 285
[hep-ph/0105306].

\bibitem{Dimopoulos:1981zb}
S.~Dimopoulos and H.~Georgi,
Nucl.\ Phys.\ B {\bf 193} (1981) 150.

\bibitem{Sakai:1981gr}
N.~Sakai,
Z.\ Phys.\ C {\bf 11} (1981) 153.

\bibitem{Ellis:1981ts}
J.~R.~Ellis and D.~V.~Nanopoulos,
Phys.\ Lett.\ B {\bf 110} (1982) 44.

\bibitem{Donoghue:1983mx}
J.~F.~Donoghue, H.~P.~Nilles and D.~Wyler,
Phys.\ Lett.\ B {\bf 128} (1983) 55.

\bibitem{Giudice:1998bp}
G.~F.~Giudice and R.~Rattazzi,
Phys.\ Rept.\  {\bf 322} (1999) 419
[hep-ph/9801271].

\end{thebibliography}
\end{document}